# Electric-Field-Tunable Luttinger compensated antiferromagnetism in double CrCl$_2$ chains

Deping Guo[1,*], Weihan Zhang[2,3], Canbo Zong[2,3], Cong Wang[2,3] and Wei Ji[2,3]

[1]*College of Physics and Electronic Engineering, Center for Computational Sciences, Sichuan Normal University, Chengdu, 610101, China*

[2]*Beijing Key Laboratory of Optoelectronic Functional Materials & Micro-Nano Devices, School of Physics, Renmin University of China, Beijing 100872, China*

[3] *Key Laboratory of Quantum State Construction and Manipulation (Ministry of Education), Renmin University of China, Beijing, 100872, China*

[*]*Corresponding author.* Email: dpguo@sicnu.edu.cn (D.G.)

**Abstract**

Luttinger compensated antiferromagnets (LcAFMs), combining spin polarization with vanishing net magnetization, offering distinct advantages for next-generation spintronic applications. Using first-principles calculations, we demonstrate that conventional antiferromagnetic CrCl$_2$ double chains can be transformed into one-dimensional LcAFMs under an external electric field, exhibiting pronounced isotropic spin splitting. The magnitude of the splitting, as well as the band gap, can be effectively tuned by both in-plane and out-of-plane fields, thereby providing greater controllability than in two-dimensional counterparts. To further enhance the tunability, we design a nearly lattice-matched CrCl$_2$/MoTe$_2$ heterostructure and uncover that interfacial charge transfer generates a built-in electric field, inducing spin splitting comparable to that driven by external fields. These results establish interfacial engineering as a highly efficient route to realize and manipulate LcAFM states in low-dimensional magnets, expanding the design principles for spintronic functionalities at the nanoscale.

Conventional antiferromagnets (AFM) are appealing platforms for ultrahigh-density spintronic integration because their negligible stray fields suppress magnetic crosstalk [1]. However, the absence of net magnetization hinders conventional routes to spin-polarized charge currents and complicates electrical readout, although spin-orbit and interfacial effects can partially alleviate these limitations. Recently, altermagnets have been identified as collinear magnets that retain zero net magnetization yet host momentum-dependent spin polarization without requiring relativistic spin–orbit coupling [2,3]. The spin polarization arises from spin group symmetries that break the joint parity-time (P-T) or time-reversal-fraction-translation (T$\tau$) symmetry. When these symmetries are further lifted, the altermagnetic state evolves into a Luttinger-compensated antiferromagnet (LcAFM), which exhibits isotropic spin splitting [4–10] across the entire Brillouin zone (BZ). LcAFMs remain collinear and magnetically compensated but exhibit spin polarization [4–6,11] that display anomalous Hall and magneto-optical responses even in the absence of spin–orbit coupling [4]. The exact magnetic compensation in LcAFMs is not protected by symmetry but arises from electron filling, which ensures an exact balance between spin-up and spin-down occupations despite the presence of localized magnetic moments [4–8,12–22]. Because the opposite-spin sublattices are not symmetry-related, LcAFMs can be robust against external perturbations.

Extending the family of LcAFMs from three dimensions into low-dimensional materials opens promising avenues for tunable quantum functionalities and integration into nanoscale spintronic devices. Numerous altermagnets have been theoretically proposed and some experimentally verified [23,24], the number of predicted LcAFMs in bulk materials remains limited [5,8,18,25,26], and even fewer candidates have been predicted in two dimensions [4,9,27]. A recent theoretical study suggests that 2D LcAFMs could be realized by lifting the symmetries that relate the two spin sublattices through Janus structures, electric-field- or substrate-induced staggered potentials, or elemental substitution [4]. These strategies render the two spin sublattices symmetry-unrelated, thereby exhibiting spin splitting across entire BZ. One-dimensional (1D) single-atomic magnetic chains were recently synthesized in experiments [28–32], establishing them as an emerging member of low-dimensional magnet family. Their inter-chain interactions are governed by similar mechanisms to interlayer couplings in 2D magnets, sharing the same features of highly tunable and strongly influential on the overall magnetism. In our previous work, we theoretically demonstrated the emergence of altermagnetic states in quasi-one-dimensional layered structures self-assembled from 1D magnetic atomic chains [33]. This naturally raises a question: can assembled 1D magnetic chain structures host LcAFMs?

In this work, we theoretically explore the magnetic properties of experimentally synthesized AFM $CrCl_2$ double chains [29] under external electric fields. Using density functional theory (DFT) calculations, we reveal that the application of an electric field breaks both the inversion symmetry and $\{C_{2x}|(1/2,0,0)\}$ symmetry operation connecting opposite sublattices, thereby driving a phase transition into an LcAFM state characterized by pronounced isotropic spin splitting. The spin splitting magnitude and the band gap can be effectively tuned by the electric field strength. To realize a permanent electric field, we further construct a heterostructure of $CrCl_2$ double chains on a nearly lattice-matched $MoTe_2$ substrate, where the substrate-induced built-in field stabilizes the LcAFM states.

Our DFT calculations were carried out using the generalized gradient approximation for the exchange-correlation potential [34], the projector augmented wave method [35] and a plane-wave basis set as implemented in the Vienna ab-initio simulation package (VASP) [36,37]. In all calculations, the Grimme's D3 form vdW correction was applied to the Perdew Burke Ernzerhof (PBE) exchange functional (PBE-D3) [38]. Kinetic energy cut-offs of 700 eV and 500 eV for the plane wave basis set were used in structural relaxations and electronic calculations, respectively. The structures were fully relaxed until the residual force per atom was less than 0.001 (0.01) eV/Å for free-standing (substrate-supported) $CrCl_2$ double chains. A 22×1×1 (14×2×1) $k$-mesh was adopted to sample the Brillouin zone of free-standing (substrate-supported) $CrCl_2$ double chains. A vacuum layer, over 15 Å in thickness, was used to reduce interactions among image slabs. On-site Coulomb interactions on the Cr $d$ orbitals were considered using a DFT+U method [39] with $U$= 6.2 eV and $J$ =1.0 eV, consistent with the values used in the literature [29]. Phonon spectra were calculated using the density functional perturbation theory, as implemented in the PHONOPY code [40]. In phonon spectrum calculation, the dispersion correction was made at the van der Waals density functional (vdW-DF) level [41] with the optB86b functional for the exchange potential (optB86b-vdW) [42]. In constructing the $CrCl_2$/$MoTe_2$ heterostructure, a bilayer $MoTe_2$ substrate was adopted. During structural relaxation, the bottom layer of $MoTe_2$ was fixed, while the top layer and the $CrCl_2$ double chains were fully relaxed.

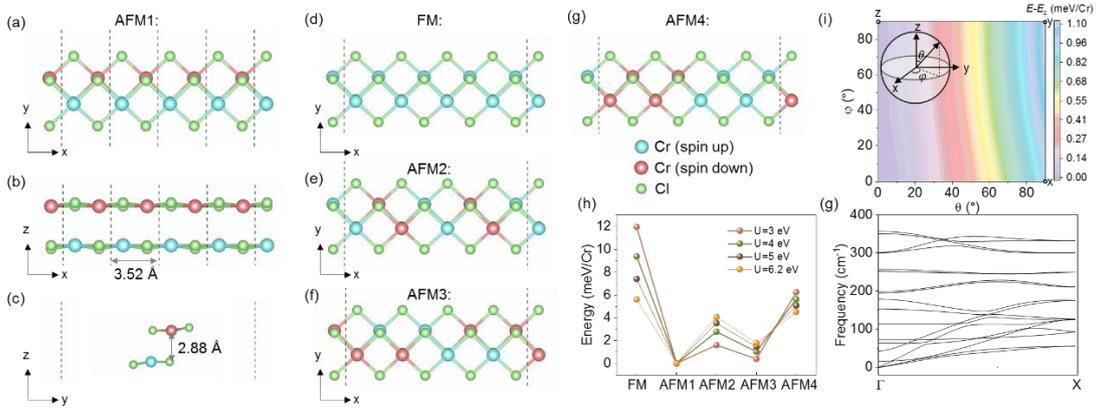

FIG. 1. (a) Top and (b)-(c) side views of $CrCl_2$ double chains under AFM1 magnetic configuration. (d)-(g) Top views of $CrCl_2$ double chains under (d) FM, (e) AFM2, (f) AFM3 and (g) AFM4 magnetic configurations. (h) Relative energies of five magnetic configurations as a function of $U$ at $J$=1.0 eV. (i) Magnetic anisotropy energy mapping of $CrCl_2$ double chains with the AFM1 configuration. The coordinate system was defined as shown. The zero energy is defined by the configuration in which the magnetic moments align parallel to the $z$-direction. (g) Phonon spectra of $CrCl_2$ double chains with the AFM1 configuration.

To realize 1D LcAFM states through the construction of double-chain structures, it is first required that the total magnetic moment of the double chain vanishes. This condition necessitates AFM ordering either within each chain or between the chains. The stable phase of 1D $CrCl_2$ is the α-phase [43]. When two $CrCl_2$ chains are combined to form a double chain, the periodic axis is oriented along the $x$-direction [Figs. 1(a)-1(b)], along which the double-chain structure was constructed and fully relaxed. The two chains adopt a staggered stacking, with Cr atoms facing the Cl atoms of the adjacent chain [Figs. 1(a)-1(c)], rather than a direct AA stacking, which exhibits imaginary phonon frequencies. Considering five magnetic configurations [Figs. 1(a)-1(g)], we find that the magnetic ground state corresponds to intrachain FM and interchain AFM (AFM1) [Fig. 1(a)]. This configuration remains energetically preferred across the range of Hubbard $U$ values considered [Fig. 1(h)]. In this magnetic ground-state configuration, the intrachain lattice constant is 3.52 Å, and the interchain distance is 2.88 Å [Figs. 1(b)-1(c)], which is larger than the Cr-Cl bond length (2.53 Å) in 2D $CrCl_2$. The easy axis of magnetization lies along the $z$-direction, with a magnetic anisotropy energy of 1.1 meV/Cr [Fig. 1(i)]. The phonon spectrum exhibits no significant imaginary frequencies [Fig. 1(g)], confirming the dynamical stability of the double chain.

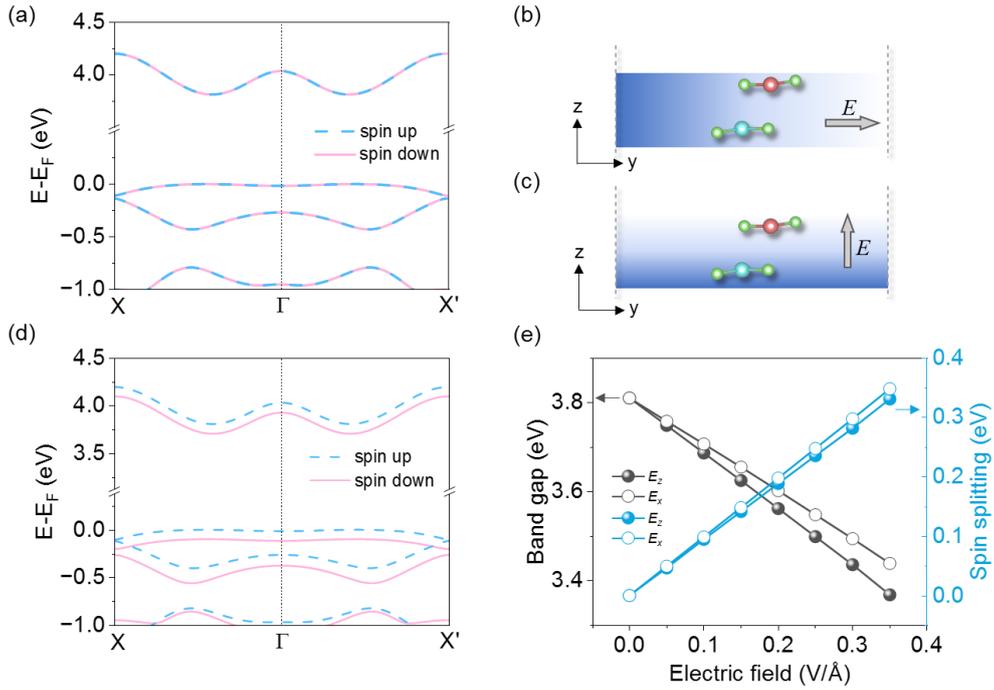

FIG. 2. (a) Band structure of $CrCl_2$ double chains under AFM1 configuration. (b)-(c) Schematic illustration of external electric field applied along the (b) $y$- and (c) $z$- directions of the $CrCl_2$ double chains. (d) Band structure of $CrCl_2$ double chains under an external electric field at $E_y$=0.1 V/Å. (e) Variation of the spin splitting at the valence band maximum ($E_{\text{spin down}}$-$E_{\text{spin up}}$) and the band gap as a function of the external electric field.

The space group of the CrCl$_2$ double chains is P2$_1$/m (No. 11). Sublattices with opposite spins are related through either inversion symmetry or the operation {$C_{2x}$|(1/2,0,0)}, thereby ensuring PT symmetry and classifying the system as a conventional AFM. From the band structure [Fig. 2(a)], it is evident that the spin-up and spin-down states remain degenerate, with a band gap of 3.81 eV. Upon the application of either an in-plane or out-of-plane electric field [Figs. 2(b)-2(c)], however, both the inversion symmetry and the {$C_{2x}$|(1/2,0,0)} operation in the CrCl$_2$ double chains are broken as a result of the electrostatic potential difference. Consequently, the sublattices with opposite spins are no longer connected through any symmetry operation, leading to the emergence of LcAFM states.

In comparison with two-dimensional (2D) LcAFM systems, the 1D LcAFM not only permits the modulation of its electronic states through an out-of-plane electric field but also allows for precise control via an in-plane field, thereby providing an additional degree of tunability. When an in-plane electric field of 0.1 V/Å is applied, an isotropic spin splitting emerges, with a magnitude of approximately 100 meV at the valence band maximum [Fig. 2(d)]. Concomitantly, the bandgap is reduced to 3.71 eV from 3.81 eV. We systematically analyzed the evolution of spin splitting and bandgap as a function of external electric fields [Fig. 2(e)]. Both in-plane and out-of-plane electric fields produce analogous trends: as the field strength increases, the spin splitting progressively enlarges. Under the applied field, the conduction band minimum (spin-down, pink solid line) shifts downward, while the valence band maximum (spin-up, blue dashed line) shifts upward (Fig. S1). Consequently, with increasing field strength, the bandgap of the CrCl$_2$ double chains diminishes monotonically [Fig. 2(e)].

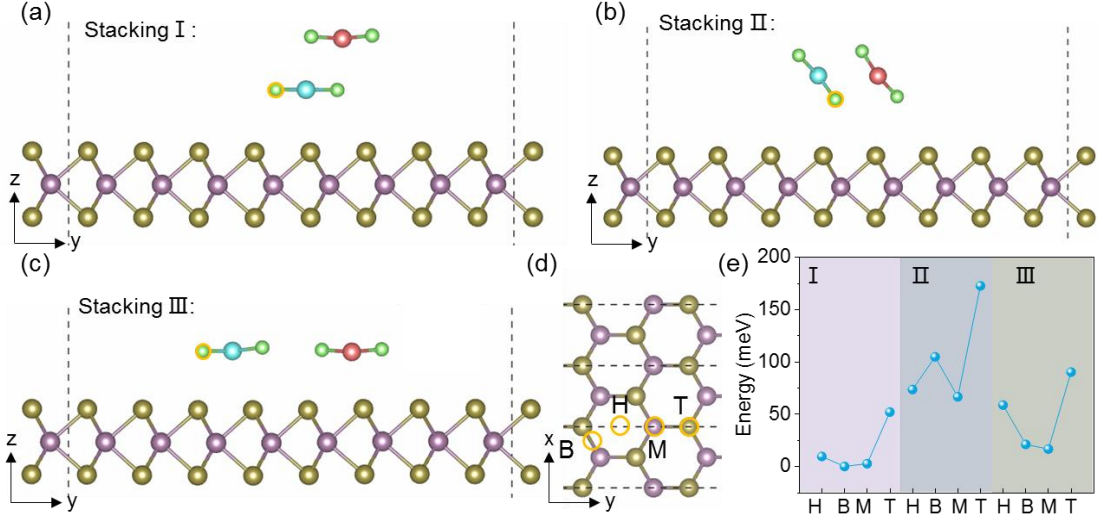

FIG. 3. (a–c) Side views of three stacking styles of the $CrCl_2$/$MoTe_2$ heterostructure: (a) vertical (I), (b) tilted-parallel (II), and (c) parallel orientations (III). A freestanding bilayer $MoTe_2$ is used in these structures with only the top layer displayed. (d) Considered four initial configurations for the Cl atom [highlighted by yellow circles in panels (a–c)], including bridge (B), hollow (H), Mo-top (M), and Te-top (T) sites. (e) Relative total energies of different initial configurations under different stackings.

Charge transfer from the substrate can induce an intrinsic electric field, thereby offering a direct and more energy-efficient strategy for modulating the electronic structure of $CrCl_2$ double chains than the application of an external electric field. Accordingly, we designed a heterostructure with bilayer $MoTe_2$ as the substrate, selected for its lattice constant (a = 3.52 Å) commensurate with that of $CrCl_2$ (a = 3.52 Å), which minimizes potential strain effects on the electronic properties of the chains and renders experimental fabrication more feasible. The periodic axis (*x*-axis) of the $CrCl_2$ chains is oriented along the Zig-zag (*x*-axis) direction of $MoTe_2$ (1×1×1 $CrCl_2$ chains on $1\times4\sqrt{3}\times1$ $MoTe_2$). Taking into account the structural characteristics of the $CrCl_2$ double chains, three possible stacking styles were considered: perpendicular [stacking I, Fig. 3(a)], tilted-parallel [stacking II, Fig. 3(b)], and parallel [stacking III, Fig. 3(c)]. Each stacking style gives rise to distinct interfacial interactions between the $CrCl_2$ double chains and the $MoTe_2$ substrate. On this basis, four potential initial configurations were further evaluated for each stacking style, referenced to the position of a specific Cl atom [yellow circles in Figs. 3(a)–3(c)]: bridge, hollow, Mo-top, and Te-top sites [Fig. 3(d)]. Comparative total-energy analyses demonstrate that the most energetically favorable configuration is the fully relaxed bridge configuration in the perpendicular stacking style [Fig. 3(e) and Fig. S2]. In the most stable configuration, the strong interchain interactions within the $CrCl_2$ double chains are retained, as compared with stacking III, together with robust interfacial coupling to the substrate, in contrast to stacking II.

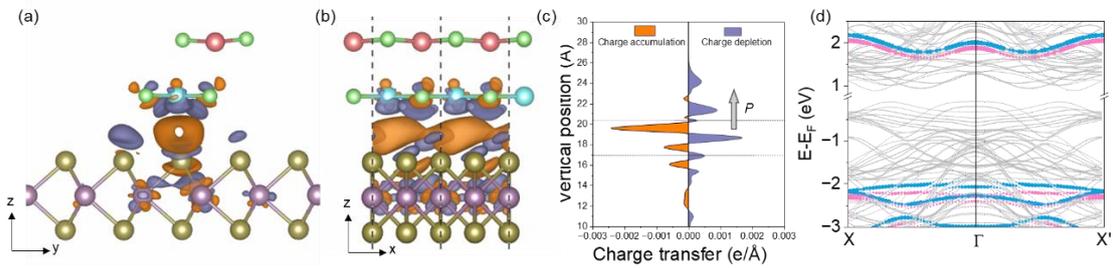

FIG. 4. (a–b) Side views of the differential charge density (DCD) at the $CrCl_2$/$MoTe_2$ interface in the most stable configuration. Orange (purple) isosurfaces represent charge accumulation (reduction). (c) The line profile of DCD along $z$ direction. The gray arrow denotes the polarization induced by interfacial charge transfer. The two dashed lines indicate the highest position of Te atoms and the lowest position of Cl atoms, respectively. (d) Projected band structure of $CrCl_2$/$MoTe_2$ heterostructure in the most stable configuration. Blue and pink bands correspond to spin-up and spin-down states of $CrCl_2$, while gray bands mainly originate from $MoTe_2$.

On the $MoTe_2$ substrate, the $CrCl_2$ double chains retain intrachain FM and interchain AFM coupling, consistent with their free-standing counterpart (Fig. S3). As shown in Fig. 4(a), the interfacial differential charge density (DCD) at the vertically stacked double-chain $CrCl_2$/$MoTe_2$ interface exhibits pronounced charge variation at the gap between the $MoTe_2$ substrate and the bottom $CrCl_2$ chain [Figs. 4(a)-4(b)], exhibiting significant charge accumulation within the gap. The interfacial charge variation induces an out-of-plane polarization [Fig. 4(c)], thereby breaking the PT and rotation symmetry in a manner analogous to applied electric fields and consequently stabilizing the LcAFM state. The projected band structure of the heterostructure further reveals a definite spin splitting of approximately 120 meV at the valence band edge spanning the whole BZ [Fig. 4(d)]. Comparable isotropic spin splittings are also observed for other stacking styles (Fig. S4), although their magnitudes differ due to the stacking-dependent effective polarization induced by the varying interfacial interactions. These results highlight the efficacy of interfacial charge transfer as a robust and versatile strategy for modulating the LcAFM state.

In summary, our study reveals that $CrCl_2$ double chains provide a robust 1D platform for realizing field-tunable LcAFM. Unlike 2D LcAFMs, the 1D LcAFM $CrCl_2$ double chains enable the breaking of PT and $\{C_{2x}|(1/2,0,0)\}$ symmetries under either in-plane or out-of-plane electric fields, thereby inducing sizable isotropic spin splitting. Meanwhile, the built-in electric field induced by interfacial charge transfer in $CrCl_2/MoTe_2$ heterostructure offers an alternative, energy-efficient pathway to achieve the same effect. This dual strategy, external field control and substrate-induced polarization, not only enriches the fundamental understanding of compensated magnetism in reduced dimensions, but can also be extended to other thermodynamically and dynamically stable 1D antiferromagnetic nanoribbons. The demonstrated tunability and versatility highlight the potential of 1D LcAFMs as promising building blocks for future spintronic technologies.

**Acknowledgements**

We gratefully acknowledge the financial support from the National Natural Science Foundation of China (Grants No. 92477205 and No. 52461160327), the National Key R&D Program of China (Grant No. 2023YFA1406500). Calculations were performed at the Hefei Advanced Computing Center, the Physics Lab of High-Performance Computing (PLHPC) and the Public Computing Cloud (PCC) of Renmin University of China.